\documentclass[aps,twocolumn,prd,showpacs,nofootinbib]{revtex4}
\usepackage{amsmath}
\usepackage{graphicx}
\usepackage{dcolumn}
\usepackage{bm}
\usepackage{amssymb}
\usepackage{latexsym}

\def\be{\begin{equation}}
\def\ee{\end{equation}}
\def\ba{\begin{eqnarray}}
\def\ea{\end{eqnarray}}

\begin{document}

\title{Scalar Perturbation Spectra
in an Emergent Cosmological Island}

\author{Yun-Song Piao$^{a,b}$}
\affiliation{${}^a$College of Physical Sciences, Graduate School
of Chinese Academy of Sciences, YuQuan Road 19{\rm A}, Beijing
100049, China} \affiliation{${}^b$Interdisciplinary Center of
Theoretical Studies, Chinese Academy of Sciences, P.O. Box 2735,
Beijing 100080, China}

\begin{abstract}

The possibility that the universe may have a fundamental and
positive cosmological constant has motivated an interesting
cosmological model, in which initially the universe is in a
cosmological constant sea, then the local quantum fluctuations
violating the null energy condition create some islands with
matter and radiation, which under certain conditions might
corresponds to our observable universe. We in this note study the
perturbation spectra of scalar fields not affecting the evolution
of background during the fluctuation.
We will examine whether they can be
interesting, and responsible for the structure formation of
observable universe.

\end{abstract}

 \maketitle

The inflation is expected to take place at the earlier moment of
the universe, which superluminally stretched a tiny patch of space
volume to our observable universe. This not only
phenomenologically successfully predicts a nearly scale-invariant
spectrum of adiabatic cosmological perturbations, but can also be
motivated very reasonably by many scalar fields naturally arising
from string theory \cite{Lin, JC}. However, it is well-known the
inflation has some inevitable fine-tune problems, in addition also
suffered from some conceptual embarrassments, such as the initial
singularity, the cosmological constant, as has been described in
Ref. \cite{B}. Thus though how embedding the inflation scenario
into a realistic high-energy physical theory has received more and
more attentions, it might be still interesting or desirable to
search for an alternative to the inflation scenario \cite{HW, W}.

Recent observations imply that the universe may have a fundamental
and positive cosmological constant. This result means that the
universe will generally asymptotically approach everlasting dS
equilibrium state. Thus it may be expected that the most unlikely
process, such as the creation of local universes, will inevitably
occur \cite{DKS}. This in some sense has motivated an interesting
cosmological model
\cite{DV, P}, in which initially the universe is in a cosmological
constant ($\Lambda$) sea, then the local quantum fluctuations
violating the null energy condition (NEC) will bring some
explosive events and create some islands with matter and
radiation, which under certain conditions might correspond to our
observable universe. During the fluctuation the Hubble length
scale decreases rapidly, which makes the perturbations have an
opportunity to leave the horizon and become the primordial
perturbations seeding the structure of our observable universe
when their reentering the horizon during radiation/matter
domination. However, instead of undergoing a contraction in the
Einstein frame, as in Pre Big Bang (PBB) \cite{VGV, V} and
ekpyrotic/cyclic scenario \cite{KOS} in which the primordial
perturbations are generated during this contracting phase, in the
island universe model the local scale factor will expand all the
time during and after the NEC violating fluctuation. Thus in this
sense here the thermalization after the fluctuation actually means
a transition from one expanding phase to another.
The island universe model is to some extent similar to the
recycling universe proposed by Garriga and Vilenkin \cite{GV}. The
difference is that in the former the fluctuation is not followed
by an inflation, instead the thermalization will occur instantly
and the radiation fills the volume rapidly.

Whether the island universe model is consistent with the
observations has been discussed in Ref. \cite{P}, in which the NEC
violating fluctuation was simulated by a scalar field violating
the NEC, see also \cite{DV, D} for a different study. This in some
sense corresponds to introduce a creation field, as in steady
state cosmology \cite{BG, H, HN}.
The primordial perturbation of the NEC violating field has been
studied in Ref. \cite{PZ, PZ1}, in which we firstly showed that
when the parameter $\epsilon$, which describes the abrupt degree
of background change and will be seen in the following, approaches
the negative infinity, the nearly scale-invariant spectrum of
adiabatic fluctuations can be obtained, which might be inherited
by late-time radiation/matter-dominated phase. However, the result
is severely dependent of whether the growing mode of spectrum of
Bardeen potential $\Phi$ before the transition can be transferred
to the constant mode after the transition, which is similar to the
case of the ekpyrotic/cyclic scenario \cite{DurrerV, CDC, GKST,
PZ2}. Thus there is an uncertainty around the physics of
transition from the NEC violating fluctuation to the
radiation-dominated phase. In the simple and conventional scenario
it seems that the growing mode of $\Phi$ can hardly be matched to
the constant model after the transition \cite{BF, Lyth, Hwang, T,
TBF}, which has been also further shown by some numerical studies
\cite{C, AW, BV}. Recently, however, it has been illuminated
\cite{Bozza} that whether the final spectrum is that of the
growing mode before the transition depends on the presence or
absence of the direct relation between the comoving pressure
perturbation and $\Phi$ in the energy momentum tensor, in which
the new physics mastering the transition might be encoded. This is
also in agreement with the earlier result of Ref. \cite{DurrerV}.
However, in this note instead of that of the NEC violating
background, we
will study the perturbation spectra of other scalar fields
possibly appearing in the island universe. We assume that these
scalar fields do not affect the evolution of background during the
fluctuation. We will examine whether their appearance will make
the background after the fluctuation become so inhomogeneous as to
be unaccepted, and in what case they can generate the scale
invariant spectra and the valid amplitude of perturbations
responsible for the structure formation.

We firstly briefly show some model-independent characters of
island universe. The NEC violating fluctuations in the $\Lambda$
sea, in which $\Lambda$ is the present value of cosmological
constant, will create some thermalized matter islands, which
subsequently evolute as the usual FRW universe. The radiation and
matter will be diluted with the expansion of island and eventually
this part of volume will return to the $\Lambda$ sea again. The
total evolution may be visualized in Fig. 1. We set $8\pi/m_p^2
=1$ and work with the parameter $\epsilon \equiv -{\dot h}/h^2$,
where $h\equiv {\dot a}/a$ is the local Hubble parameter. The
``local" means here that the quantities, such as $a$ and $h$, only
character the values of the NEC violating region. The scale of the
NEC violating region is required to be larger than the horizon
scale of $\Lambda$ sea \cite{VT, DV}, which is consistent with
Refs. \cite{FG, Linde92} and in some sense suggested by the
singularity theorems, see also \cite{AJ}. This corresponds to set
the initial values of the relevant quantities. The $\epsilon$ can
be rewritten as $\epsilon \simeq {1\over h \Delta t}{\Delta h\over
h}$, thus in some sense $\epsilon$ actually describes the change
of $h$ in unit of Hubble time and depicts the abrupt degree of
background fluctuation.
The reasonable and simplest selection for the local evolution of
scale factor $a(t)$, due to ${\dot h}
>0 $, is \cite{PZ, PZ2} $ a(t) \sim (-t)^{n}$, where $t$ varies from
$-\infty$ to $0_-$, and $n$
is a negative constant. 
Thus $\epsilon\equiv 1/n$. In the conformal time, we obtain \be
a(\eta) \sim (-\eta)^{{n\over 1-n}}\equiv (-\eta)^{1\over \epsilon
-1} . \label{aeta}\ee Then we can derive \be a\sim \left({1\over
(1-\epsilon) a h}\right)^{1\over \epsilon -1}\equiv \left({n\over
(n- 1) a h}\right)^{n\over 1- n} .\label{ah}\ee To make the time
of the NEC violating fluctuation $T=\int dt \rightarrow 0$, for
which the fluctuate can be so strong as to be able to create some
islands with the enough thermalization temperature \cite{DV},
$|\epsilon | \gg 1$ is required \cite{P}, which results in
$n\rightarrow 0_-$.
The efolding number which measures
the period of the NEC violating fluctuation can be defined as \be
{\cal N}_{ei} \equiv \ln{({a_e h_e\over a_i h_i})}
,\label{caln}\ee where the subscript `i' and `e' denote the values
at the beginning and end of the fluctuation, respectively. In the
island universe model $h_i^2 \equiv h_0^2\sim \Lambda$, where the
subscript 0 denotes the present value. Thus from (\ref{ah}), we
have \be {a_e\over a_i}=({a_i h_i\over a_e h_e})^{1\over \epsilon
-1}= e^{{{\cal N}\over 1-\epsilon}} .\ee We can see that for the
negative enough $\epsilon$, the change $\Delta
a/a=(a_e-a_i)/a_i\simeq {\cal N}/(1-\epsilon)$ of $a$ will be very
small. Thus from (\ref{caln}) and taking $a_e\simeq a_i$, we can
approximate the efolding number as ${\cal N}_{ei} \simeq \ln{(h_e/
h_i)}$. Further, it has been shown \cite{P} that as long as the
islands were created, the efolding number required to solve the
horizon problem of FRW universe may be always enough, which is
independent of the energy scale of thermalization.

\begin{figure}[t]
\begin{center}
\includegraphics[width=8cm]{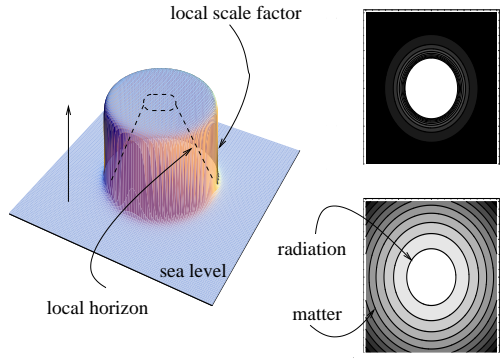}
\caption{The left figure is the sketch of an emergent island
universe in $\Lambda$ sea, which arises from a local quantum
fluctuation violating the null energy condition, in which during
the fluctuation the local Hubble length scale decreases rapidly,
but the local scale factor will expand all the time, though its
change is so small as to be nearly unchanged. The upper right
panel is the contour plot of left figure, in which the white
region is the island filled with radiation, while the black region
is the $\Lambda$ sea. The lower right panel look likes the ripples
on the sea, which is the contour plot of left figure after an
enough long time since the fluctuation is over, in which the
internal and external rings represent radiation/matter-dominated
region, respectively. There is not the inflation (slow rolling)
region, compared with the recycling universe \cite{GV}. }
\end{center}
\end{figure}

We will discuss the perturbation spectra of scalar field $\varphi$
not affecting the evolution of the background in the following. In
the momentum space, the motion equation of $\varphi$ can be
generally written as \be u_k^{\prime\prime}+(k^2-f(\eta))u_k =0
,\label{uk}\ee where $u=a\varphi$ and $ f(\eta)= (v^2 -1/4)/
\eta^2$, $v$ is determined by the details of $\varphi$ field, such
as its mass, its couple to the background, as will be seen in the
following. The general solutions of this equation are the Hankel
functions. In the regime $k\eta \rightarrow \infty $, the mode
$u_k$ is very deep into the horizon of $\Lambda$, in which $u_k$
can be well defined and taken as $ u_k \sim {e^{-ik\eta} \over
\sqrt{2k}}$, which sets the initial condition of its evolution. In
the regime $k\eta \rightarrow 0$, the mode $u_k$ is far out the
horizon, and become unstable and grows. In long wave limit, we can
expand the Hankel functions to the leading term of $k$ and obtain
\be k^{3/2}\varphi_k \simeq ({n-1\over n})^{v-1/2} ({h\over
2\pi})({k\over ah})^{3/2-v} ,\label{phik}\ee where the coefficient
factor of ${\cal O}(1)$ has been neglected, and $\eta
=(n/(n-1))(-1/ah)$ has been used. Thus we see that if $v={3\over
2}$, the spectrum of $\varphi$ field will be scale-invariant.
Further since the local Hubble parameter $h$ of the background is
rapidly changed during the NEC violating fluctuation, the
amplitude of perturbations of $\varphi$ will continue to increase
after its leaving the horizon, up to the end of fluctuation. Then
the perturbations of $\varphi$ will be transferred to the
curvature perturbations in the radiation, as will be discussed in
the third last paragraph, and thus become a constant in the
superhorizon scale. Since the perturbations of $\varphi$ is
increased in the superhorizon scale during the NEC violating
fluctuation and is constant after the end of fluctuation, it can
be reasonable to take the value at the time when the NEC violating
fluctuation is over to calculate the amplitude of perturbations,
\be k^3|\varphi_k|^2 \simeq ({n-1\over n})^2 ({h_e\over 2\pi})^2
.\label{phik2}\ee We can see that for the inflation model in which
$n\rightarrow \infty$, the factor $({n-1\over n})^2\simeq {\cal
O}(1)$ of (\ref{phik2}) can be generally neglected, thus the
perturbation amplitude of ``test" scalar field with
scale-invariant spectrum is generally $(h/2\pi)^2$, which is
well-known result for the inflation models. However, in the island
universe model since $n\equiv 1/\epsilon \simeq 0_-$, there is a
significant magnification of amplitude $\sim ({n-1\over
n})^2=(1-\epsilon)^2 \sim |\epsilon|^2$, which can be larger with
that $\sim |\epsilon|$ of Bardeen potential \cite{P} \be
k^3|\Phi_k|^2\simeq {1\over n} ({h_e\over 2\pi})^2
.\label{Phik}\ee Thus the primordial perturbation of ``test" field
$\varphi$ can be interesting for further studies.

There have been still some doubts that the island universe model
can work, since intuitively the fluctuation occurs more possibly
in a smaller scale, which implies that the fluctuation with the
cosmological scale should be rather inhomogeneous.
However, Eqs. (\ref{phik2}) and (\ref{Phik}) means that the
homogeneousness of fluctuation is to a great extent dependent of
the abruptness $\epsilon$ of background change and the energy
density $h_e^2$ after the fluctuation, not its distance scale,
i.e. if $\epsilon$ and $h_e$ is very large, the amplitude of
perturbations will $> {\cal O}(1)$, which will make the background
after the thermalization become very inhomogeneous and thus be
unaccepted. For example, for (\ref{Phik}), when $\epsilon>
10^{120}$ is taken, the fluctuations with arbitrary value of $h_e$
will $> {\cal O}(1)$ or even $\gg {\cal O}(1)$, and thus the
background after the fluctuation will be extremely inhomogeneous.
But if we take $\epsilon$ and $h_e$ with seemly value, we can find
that the background after the thermalization not only is very
homogeneous, but in some sense can be responsible for the
structure formation of observable universe. For example, for
(\ref{Phik}), having taken $\rho_{r}\sim m_{ew}^4 \sim 10^{-60}$
as the reheating scale, where $m_{ew}$ is the electroweak scale,
and $|\epsilon |\sim 10^{50}$, we have $k^3|\Phi_k|^2 \sim
10^{-10}$ \cite{P}, which is just the observed amplitude of CMB.
The value of $\epsilon$ reflects the abrupt degree of background
change, which in present case can hardly be determined by
underlying theory. We admit that the proper amplitude of
perturbation can be obtained only for moderate value of
$\epsilon$, while for $|\epsilon | \rightarrow \infty$, the
perturbations of both Bardeen potential and ``test" scalar field
will be divergent. Thus for our consideration, the value of
$\epsilon$ seems to require some fine-tuning and, however, might
be still anthropically accepted.

The amplitude $\sim |\epsilon|^2$ of scale-invariant spectrum of
``test" scalar field
is far larger than that of $\Phi$, which, compared with $\Phi$,
may be more possibly responsible for structure formation of
observable universe. How to obtain the scale-invariant spectrum of
$\varphi$?
We firstly regard $\varphi$ as a massless field with minimal
coupling. Thus \be f(\eta) \equiv a^{\prime\prime}/ a
\label{feta}\ee in (\ref{uk}). From (\ref{aeta}), we have
$f(\eta)= (2-\epsilon)/ \eta^2(\epsilon-1)^2 $. Thus after
applying it to the island universe model, we obtain $
v=\sqrt{(2-\epsilon)/ (\epsilon-1)^2+1/4}\simeq \sqrt{{\cal
O}(1/\epsilon)+1/4}\simeq 1/2$. Thus to the leading order of $k$,
we have $k^{3/2}\varphi_k \sim k$, which means that instead of the
scale-invariance in the inflation model, the spectrum of a very
light scalar field in the island universe model is $\sim k^2$ and
thus its amplitude will be strong suppressed at large scale. We
then turn to a massive scalar field with the constant mass
$m_{\varphi}$. The term $-m^2_{\varphi}a^2$ will be required to
add to (\ref{feta}). Thus \ba v &=& \sqrt{(2-\epsilon)/
(\epsilon-1)^2+1/4- (m_{\varphi}/h(1-\epsilon))^2}\nonumber\\ &
\simeq & \sqrt{1/4-(m_{\varphi}/|\epsilon|h)^2} .\label{vmass}\ea
However, since the $h$ changes very quickly with the time during
the perturbations are produced, the (\ref{uk}) will be not the
exact Bessel equation any more, which makes us very difficult to
obtain its analytic solution. However, for our purpose the
qualitative analysis may be enough. There are two limits, which
are $m_{\varphi}\ll |\epsilon | h$ and $m_{\varphi}\gg |\epsilon |
h$, for the arbitrary values of $h$ during the NEC violating
fluctuation, respectively. The former corresponds the massless
case. For the latter, (\ref{vmass}) will become imaginary, and
thus can not attach $v$ to the spectrum index of $\varphi_k$,
since to obtain $k^3|\varphi_k|^2$ we need to take the module
square of (\ref{phik}). This means that the spectrum $\sim k^3$
and thus will falls down very rapidly at large scale. In addition
it can be importantly, as is seen from (\ref{phik}), notice that
when $v\simeq 1/2$ or is imaginary, there is not a magnification
of amplitude from $\epsilon$. Thus for the general massless and
massive scalar field, its perturbation can be negligible during
the NEC violating fluctuation.

However, there seems to be some exceptions. Rechecking
(\ref{vmass}), one can directly find that if
$(m_{\varphi}/|\epsilon|h)^2 \simeq -2$, $v\simeq 3/2$ i.e. the
scale-invariant spectrum will be obtained, which corresponds to
introduce a time-varying mass term $m_{\varphi}^2 a^2 \simeq
-2(|\epsilon|ah)^2 = -2/\eta^2 $. This term in fact can be
motivated by a non-minimally coupling to gravitation, $\sim
R\varphi^2$, where $R\sim 1/\eta^2$ is the Ricci curvature scalar.
However, it seems not to be satisfied that this introduced term is
a negative mass term, and very large when $\eta\rightarrow 0_-$,
thus will bring a tachyonic instability. However, since initially
we may place the field $\varphi$ in $\varphi\simeq 0$, which can
be ensured by adding a small and time independent mass term $
R\sim h_0^2<m_{\rm add}^2\ll m_{\varphi}^2$ to $m_{\varphi}^2$,
and also note that the time of the NEC violating fluctuation is
actually very short $\sim {1/|\epsilon|}$, the effect of this
instability on the evolution of background might be neglected.

We also may regarded $\varphi$ as a non-normal massless scalar
field, i.e. consider a coupling between its kinetic term and
background. We adopt the work hypothesis for simplification, as in
Ref. \cite{P}, in which the NEC violating fluctuation is carried
out by a scale field $\chi$ with the reverse sign in its dynamical
term, in which the NEC $\rho +p =-{\dot \chi}^2 <0 $ is violated.
We write the coupling between the kinetic term of ``test" scalar
field $\varphi$ and the background as $g(\chi)
(\partial_{\mu}\varphi)^2$, where $g(\chi)=e^{\alpha\chi}/2$. We
have $e^{-\sqrt{-2\epsilon}\chi}\sim 1/(-t)^{2} $ \cite{PZ}, which
corresponds to a very steep potential. Thus after combining it and
(\ref{aeta}), we obtain $ e^{\chi}\sim
(-\eta)^{\sqrt{-2\epsilon}/(\epsilon-1)} $. In the momentum space,
the motion equation of $\varphi$ has the same form as (\ref{uk}),
but instead of $u=a\varphi$, here $u=b\varphi$ and $f(\eta)$ is
also changed as \be f(\eta)= b^{\prime\prime}/b ,\ee where \be
b\equiv e^{\alpha\chi/2} a \sim (-\eta)^{(\alpha \sqrt{-\epsilon/2
}+1)/(\epsilon -1)} .\ee Thus $ v=|{\alpha \sqrt{-\epsilon/2
}+1\over \epsilon -1} +1/ 2 |$. To make $v=3/2$, it seems that
$\alpha =- \sqrt{-2\epsilon}$ or $(2-3/\epsilon)
\sqrt{-2\epsilon}\simeq 2\sqrt{-2\epsilon}$ will be required,
which corresponds to a very steep coupling between $\varphi$ and
the background. These relevant discussions can be seen earlier in
the studies of PBB scenario \cite{V}.

We have found that it is still possible to generate the
scale-invariant spectrum of $\varphi$ perturbation, though the
required conditions look like very special. The reason is that the
scale-invariance of spectrum generally requires $f(\eta)\simeq
2/\eta^2$, in which $v\simeq \sqrt{2+1/4}= 3/2$, while in the
island universe model for a massless scalar field $f(\eta)\simeq
(1/\epsilon) (1/\eta)^2 \sim 0_-(1/\eta)^2 $. Thus the key
obtaining the scale-invariant spectrum is reduced to induce a
time-varying mass term $\simeq 2/\eta^2$ in $f(\eta)$, which can
not be done easily. Thus it seems that we have to introduce the
coupling of $\varphi$ to the background, either in its mass or
potential term, or in its kinetic term to acquire it.

To make the vacuum fluctuations of $\varphi$ be promoted at the
time of their leaving horizon and become classical perturbations
responsible for the structure formation of observable universe, we
need to transfer the perturbations of $\varphi$ to the curvature
perturbations in the radiation after the thermalization. This can
be accomplished when the field $\varphi$ determines the coupling
constant or the mass relevant with the decay rate of the decaying
particle during the reheating and thermalization \cite{DGZK}, see
also \cite{DGZ, Tsu, BCK, KRV, BGS}. Taking the coupling as
$\lambda (1+\varphi/m_*)$, one can have \cite{DGZK} \be {\delta
\rho \over \rho} \sim {\delta T_r \over T_r}\sim {\delta
\Gamma\over \Gamma}\sim {\varphi \over m_*} ,\label{derho}\ee
where $T_r$ is the reheating temperature. Thus combining
(\ref{phik2}) and (\ref{derho}), we have \be ({\delta \rho \over
\rho})^2\sim {1\over m_*^2}({n-1\over n})^2 ({h_e\over 2\pi })^2 .
\label{drho2}\ee Thus it seems to be very simple for (\ref{drho2})
to reproduce the observed value of the primordial perturbation
amplitude. For example, we can take $\epsilon=1/n\sim 10^{20}$,
$h_e^2\simeq m_{ew}^4\sim 10^{-60}$ and $m_*\sim 10^{-5}$ and
obtain the required amplitude $(\delta \rho / \rho)^2\sim
10^{-10}$.

In summary and discussion, intuitively one think that the
fluctuation with the cosmological scale looks like rather
inhomogeneous, but our calculations show that the inhomogeneous is
to a great extent dependent of the abruptness degree of background
change, and for moderate values of parameters the universe after
the thermalization can be very homogeneous. We also found that the
perturbation amplitude of scalar fields not affecting the
evolution of NEC violating background is far larger than that of
Bardeen potential, and thus can be more possibly responsible for
the structure formation of observable universe. Our result is
different from that of Ref. \cite{DV}. In their scenario, dS sea
phase is suddenly matched to a radiation-dominated phase, which
leads to \be k^3|\varphi_k|^2 \simeq ({h_{0}\over 2\pi})^2,
\label{vklambda}\ee thus the amplitude of density perturbation
calculated is too small to fit observation. The sudden matching
corresponds to neglect the NEC violating mediated phase, which
means that we can obtain (\ref{vklambda}) by only taking the
parameters of dS phase, i.e. $n\rightarrow \infty$, $v = 3/2$ and
$h=h_0$ in Eq. (\ref{phik}). However, in our calculations, we
mainly focus on the NEC violating phase, which corresponds to take
$n\simeq 0_-$ in Eq. (\ref{phik}). We can see from the result
(\ref{phik2}) obtained that it is this NEC violating phase that
magnifies the small perturbations inside dS sea. Note that the
value of $n$ is very different for the NEC violating phase
($n\simeq 0_-$) and dS phase ($n\rightarrow \infty$), thus it
seems that the infinitely sharp transition approximation
neglecting the NEC violating phase and the method focusing the NEC
violating phase will naturally lead to different results.
This may be the main reason of
discrepancy between Ref. \cite{DV} and us.

The perturbation of ``test" scalar field is independent of the
matching details between the NEC violating phase and
radiation-dominated phase, which to some extent relaxes the
uncertainty of perturbation spectrum of Bardeen potential
resulting from the loophole of its propagating through the
thermalization surface.
Thus the relevant disputation how the Bardeen potential propagates
through the thermalization surface is not important here. This is
very similar to the case of curvaton in inflation. The difference
is that since the expansion during inflation is nearly
exponential, to obtain a scale-invariant spectrum the curvaton is
required to be nearly massless, while since the expansion during
NEC violating phase is nearly constant, to obtain a
scale-invariant spectrum the ``curvaton" here is required to be
massive or nonminimal coupling. The conditions required look like
very
special
. However, note that a lot of scalar field in superstring theory
or other high energy theory can be very prevalent, thus it may be
imagined that some of them might have desirable behaviors, which
in some sense gives us an interesting expectation that we might
live in an emergent cosmological ``island".



\textbf{Acknowledgments} This work was partly completed in
Interdisciplinary Center of Theoretical Studies, Chinese Academy
of Sciences, and is also supported in part by NNSFC under Grant
No: 10405029, 90403032 and by National Basic Research Program of
China under Grant No: 2003CB716300.

\end{document}